\documentclass[final,5p,times,twocolumn]{elsarticle}
\usepackage[colorlinks=true, pdfstartview=FitV, linkcolor=blue, citecolor=red, urlcolor=magenta]{hyperref}
\usepackage{graphicx}
\usepackage{mwe}
\usepackage{latexsym}
\usepackage{amsmath}
\usepackage{amsfonts}
\usepackage{amssymb}
\usepackage{verbatim}
\usepackage{dcolumn}
\usepackage{amssymb}
\usepackage{bm}
\usepackage{float}
\usepackage{subfigure}
\usepackage[english]{babel}
\usepackage{tikz-cd}
\newcommand{\be}{\begin{equation}}
\newcommand{\ee}{\end{equation}}
\newcommand{\bea}{\begin{eqnarray}}
\newcommand{\eea}{\end{eqnarray}}

\newcommand{\pa}{\partial}

\journal{Physics Letters B}

\begin{document}

\begin{frontmatter}

\title{Thermal dimensional reduction and black hole evaporation}

\newcommand{\Lavras}{
\address{Physics Department, Federal University of Lavras, Caixa Postal 3037, 37200-000 Lavras-MG, Brazil}
}
\newcommand{\JP}{
\address{Department of Chemistry and Physics, Federal University of Para\'iba, Rodovia BR 079 - km 12, 58397-000 Areia-PB,  Brazil}}

\newcommand{\Itajuba}{
\address{Institute of Mathematics and Computation, Federal University of Itajub\'a, 37500-903 Itajub\'a, Brazil}
}

\author{I. P. Lobo\corref{cor1}\fnref{label1}}
\ead{lobofisica@gmail.com}
\JP
\Lavras
\author{G. B. Santos}
\ead{gbsantos@unifei.edu.br}
\Itajuba
\fntext[label1]{Corresponding author}
\begin{abstract}
We analyze how a quantum-gravity-induced change in the number of thermal dimensions (through a modified dispersion relation) affects the geometry and the thermodynamics of a charged black hole. To that end we resort to Kiselev's solution as the impact such modifications have on the evaporation rate of the black hole becomes more clear. As an application, we study the case for which the thermal dimension is reduced to two.
\end{abstract}


\end{frontmatter}

%%%%%%%%%%%%%%%%%%%%%%%%%%%%%%%%%%%%%%%%%%%%%%%%%%%%%%%%%%%%%%%%%%%%%%%%%%%%%%%%%%%%%%%%%%%%%%%%%%%%%%%%%%%%%%%%%%%%%%%%%%%%%%%%%%%%%%%%%%%%%%%%%%%%%%%%%%%

\section{Introduction}

%%%%%%%%%%%%%%%%%%%%%%%%%%%%%%%%%%%%%%%%%%%%%%%%%%%%%%%%%%%%%%%%%%%%%%%%%%%%%%%%%%%%%%%%%%%%%%%%%%%%%%%%%%%%%%%%%%%%%%%%%%%%%%%%%%%%%%%%%%%%%%%%%%%%%%%%%%%
One of the few shared features of different quantum gravity scenarios seems to be a modification in the number of dimensions of the associated quantum geometries that are expected to emerge in the so called ultraviolet (UV) regime, where we no longer can appeal to the classical Riemannian picture. Indeed, the notion of Hausdorff dimension cannot be applied to a quantum space-time \cite{Wheater:1998jb,Carlip:2009kf} and a suitable new concept should be introduced. In \cite{Amelino-Camelia:2016sru} it is argued that the concept of spectral dimension that is usually used to deal with this issue (for instance on Causal Dynamical Triangulation \cite{Ambjorn:2005db}, Asymptotic Safety \cite{Litim:2003vp}, Horava-Lifshitz gravity \cite{Horava:2009if}, Causal-Sets \cite{Belenchia:2015aia}, Loop Quantum Gravity \cite{Modesto:2008jz,oriti}) presents some inadequacies and it is suggested that a new quantity based on thermal properties of photons can provide a more physically sensible notion of dimensionality in this regime (see also \cite{Husain:2013zda,Nozari:2015iba} for related discussions). This relies on the fact that the Stefan-Boltzmann law gives, for a gas of radiation in a space with $D$ dimensions, the relation $$\rho\propto T^{D+1}$$ between energy density and temperature. Other thermodynamic quantities are also sensitive to the spatial number of dimensions and this quantum-gravity induced change of dimensions can have a decisive role, in particular, on the thermodynamics of black hole solutions in the limit where the full quantum gravity setup is not needed but its effects on the propagation of massless particles can still be relevant.
That seems to be the case, in particular, in cases of Lorentz invariance violation (LIV) or Lorentz invariance deformation in approaches to the quantum-gravity phenomenology for these black-hole systems \cite{Ling:2005bp,Ling:2005bq,Li:2008gs,Amelino:2006}, where modified dispersion relations (MDRs) are taken as key ingredients in which quantum-gravity effects are encoded.
\par
In this work we consider a generalized Horava-Lifshitz scenario and the corresponding modified thermodynamics for a black hole based on the Kiselev's solution \cite{Kiselev:2002dx}. In particular, we analyze the effect of the modified dispersion relation and its related thermal dimension on the evaporation rate of the charged (in which case a modified equation of state parameter of radiation plays a role) and uncharged black holes.
\par
This letter is organized as follows. In section \ref{sec:kiselev} we review some basic concepts of Kiselev's solution that can be used to describe a charged black hole. In section \ref{sec:thermal} we review the proposal of a thermal dimension induced by Horava-Lifshitz-inspired MDRs. In section \ref{sec:thermq} we define the relevant thermodynamic quantities that will be used and that could also be useful for future investigations, exhibiting their dependence on the thermal dimension. In section \ref{sec:evap} we use the aforementioned thermodynamic quantities along with the MDR in order to reconstruct the rate of mass loss of the charged and uncharged black holes, and discuss its asymptotic behavior. Finally, we conclude in section \ref{sec:conc}
\par
For simplicity, let us assume $c=\hbar=k_B=1$.

%%%%%%%%%%%%%%%%%%%%%%%%%%%%%%%%%%%%%%%%%%%%%%%%%%%%%%%%%%%%%%%%%%%%%%%%%%%%%%%%%%%%%%%%%%%%%%%%%%%%%%%%%%%%%%%%%%%%%%%%%%%%%%%%%%%%%%%%%%%%%%%%%%%%%%%%%%%

\section{Kiselev's solution}\label{sec:kiselev}

In this approach we assume that the black hole is surrounded by matter which, after an average process over angles is performed, can be described by the stress-energy tensor
\bea
&&T^t{}_t=T^r{}_r=-\rho(r),\\
&&T^{\theta}{}_{\theta}=T^{\phi}{}_{\phi}=\frac{1}{2}(3\omega+1)\rho(r),
\eea
where the equation of state parameter $\omega$ that relates the average pressure to the energy density is usually taken as a constant.
Considering a static space-time with spherical symmetry, the metric {\em ansatz} takes the form
\be
\label{metric}
ds^2=A(r)dt^2-\frac{dr^2}{A(r)}-r^2(d\theta^2+\sin^2\theta \,d\phi^2).
\ee
Einstein equations 
$$
G_{\mu\nu}=\kappa T_{\mu\nu},
$$
(where $\kappa=8\pi G$) under some simplifying conditions \cite{Kiselev:2002dx} then yield the solution
\bea
&&\rho(r)=\frac{C}{\kappa}\frac{3\omega}{r^{3(\omega+1)}},\\
&&A(r)=1-\frac{2GM}{r}+\frac{C}{r^{3\omega+1}},
\eea
where $C$ is a constant of integration. This approach allows one to investigate black holes surrounded by a quintessential fluid in many different scenarios (see for instance \cite{Ghosh:2017cuq,Lobo:2017dib} and references therein)\footnote{Note that the definition of quintessence in Kiselev's original paper is somewhat different from the notion we commonly see in the literature given that the stress-energy tensor in this solution is not a perfect fluid. See \cite{Visser:2019brz} for a clarification on this issue.}. Also notice that considering the averaged equation of state parameter $\omega=1/3$ one can derive the Reissner-Nordstr\"{o}m metric with electric charge in length units $Q=\sqrt{C}$ (in this case $C>0$, since $\omega>0$).\footnote{In SI units we have $Q^2=q^2G/4\pi\varepsilon_0c^4$, where $q$ is the electric charge of the black hole. In this paper, we refer to $Q$ as the electric charge, for simplicity.} This will be important in what follows.

%%%%%%%%%%%%%%%%%%%%%%%%%%%%%%%%%%%%%%%%%%%%%%%%%%%%%%%%%%%%%%%%%%%%%%%%%%%%%%%%%%%%%%%%%%%%%%%%%%%%%%%%%%%%%%%%%%%%%%%%%%%%%%%%%%%%%%%%%%%%%%%%%%%%%%%%%%%

\section{Thermal dimension}\label{sec:thermal}

We consider a class of generalized Horava-Lifshitz scenarios for which the momentum-space representation of the deformed d'Alembertian reads
\be
\Omega_{\gamma_t\gamma_x}(E,p)=E^2-p^2+\ell_t^{2\gamma_t}E^{2(1+\gamma_t)}-\ell_x^{2\gamma_x}p^{2(1+\gamma_x)},\label{MDR}
\ee
where $E$ is the energy of the particle probing the space-time, $p$ is the norm of the spatial part of its 4-momentum, $\gamma_t$ and $\gamma_x$ are dimensionless parameters, $\ell_t$ and $\ell_x$ are parameters with dimension of length (usually assumed to be of the order of the Planck length).

In order to analyze the thermal dimension we start with the partition function
\be
\log\, Q_{\gamma_t\gamma_x}=-\frac{2V}{(2\pi)^3}\int dEd^3p\left[\delta(\Omega_{\gamma_t\gamma_x})\Theta(E)2 E \log\left(1-e^{-\beta E}\right) \right].
\ee
Here $\beta$ is related to the Boltzmann constant $k_{B}$ and temperature via $\beta=\frac{1}{k_{B} T}$, and the delta function $\delta(\Omega_{\gamma_t \gamma_x})$ ensures that the on-shell relation $\Omega_{\gamma_t \gamma_x}=0$ is obeyed.

From the above expression one obtains the energy density and pressure respectively as
\be
\rho_{\gamma_t\gamma_x}=-\frac{1}{V}\frac{\partial}{\partial \beta}\log Q_{\gamma_t\gamma_x}\; , \;
p_{\gamma_t\gamma_x}=\frac{1}{\beta}\frac{\partial}{\partial V}\log Q_{\gamma_t\gamma_x}.
\ee

The trans-planckian (UV/high temperature) energy density and equation of state parameter are found to be 
\begin{eqnarray}
\rho_{\gamma_t\gamma_x}\propto T^{d^T_{\gamma_t\gamma_x}},\label{rho1}\\
\omega_{\gamma_t\gamma_x}=\frac{1}{d^T_{\gamma_t\gamma_x}-1},\label{omega1}
\end{eqnarray}
where the thermal dimension reads \cite{Amelino-Camelia:2016sru}
\be\label{dT1}
d^T_{\gamma_t\gamma_x}=1+3\frac{1+\gamma_t}{1+\gamma_x}.
\ee

These calculations are performed in the context of a flat space-time. They can be considered a good approximation in our analysis for the black hole solution as long as the wavelengths associated to the thermal photons remain small compared to the scale of curvature for the geometry in (\ref{metric}). This is exactly the case in the UV regime.

%%%%%%%%%%%%%%%%%%%%%%%%%%%%%%%%%%%%%%%%%%%%%%%%%%%%%%%%%%%%%%%%%%%%%%%%%%%%%%%%%%%%%%%%%%%%%%%%%%%%%%%%%%%%%%%%%%%%%%%%%%%%%%%%%%%%%%%%%%%%%%%%%%%%%%%%%%%

\section{Thermodynamic quantities in the trans-Planckian regime}\label{sec:thermq}
From Sec.\,\ref{sec:kiselev}, we see that it is possible to reproduce the Reissner-Nordstr\"{o}m metric by assuming $\omega=1/3$, and by setting $C=Q^2$, where $Q$ is the black hole's electric charge with dimension of length.
\par
However, if the surrounding radiation fluid obeys a modified dispersion relation, i.e., if we are in a regime in which Planck-scale corrections are relevant, we should consider the induced modifications on the equation of state parameter $\omega$. In general, $\omega$ will be temperature-dependent, a subject that was recently explored by the authors in independent investigations \cite{Santos:2015sva,Lobo:2019jdz} at the perturbative level (see also \cite{Amelino:2006} for one of the first analyses of this topic at the perturbative level). In this letter, we explore the trans-Planckian regime and the effect of the induced thermal dimension (in particular dimensional reduction) on the thermal evolution of black holes.\footnote{From now on, we shall adopt the following notation $\omega_{\gamma_t\gamma_x}\doteq \omega$ and $d^T_{\gamma_t\gamma_x}\doteq d$.}
\par
Therefore, following the same notation of previous investigations, in order to keep the length dimension of the charge, we define $C=Q^{3\omega+1}$, which along with Eq.(\ref{omega1}), gives us the following metric function in the trans-Planckian regime:
\be\label{metric1}
A(r)=1-\frac{2GM}{r}+\left(\frac{Q}{r}\right)^{(d+2)/(d-1)}.
\ee
The Hawking temperature is also dimension-dependent and reads
\begin{eqnarray}\label{TH1}
T_{H,\, d}&=& -\frac{1}{4\pi}\lim_{r\rightarrow r_+}\sqrt{-\frac{g^{rr}}{g^{tt}}}\frac{1}{g^{tt}}\frac{d}{dr}g^{tt}\nonumber\\
&=&\frac{1}{4\pi}\left[\frac{1}{r_+}-\frac{3}{d-1}\frac{1}{r_+}\left(\frac{Q}{r_+}\right)^{(d+2)/(d-1)}\right],
\end{eqnarray}
where $r_+$ is the outer horizon radius, which is defined as the larger root of the equation $A(r)=0$. The related entropy can be calculated from the identification of the black hole mass with the internal energy:
\be
S=\int\frac{1}{T_{H,\, d}}\frac{dM}{dr_+}dr_+=\frac{\pi r_+^2}{G}.
\ee
Notice that it only depends on the thermal dimension in an indirect way by means of the horizon radius that reads differently depending on the dispersion relation, whereas it is still proportional to the area of the event horizon.
\par
We define the electric potential as
\be
\Phi_{d}=\frac{\pa M}{\pa Q}\Bigg|_{S}=\frac{d+2}{2(d-1)G}\left(\frac{\pi}{GS}\right)^{3/2(d-1)}Q^{3/(d-1)},
\ee
which, of course, reduces to the usual $1/r_+$ dependence when $d=4$ or $\omega=1/3$.
\par
In this way, the first law of black hole thermodynamics gains a thermal dimension dependence as 
\be
\delta M_{d}=T_{H,\, d}\delta S+\Phi_{d} \delta Q.
\ee

%%%%%%%%%%%%%%%%%%%%%%%%%%%%%%%%%%%%%%%%%%%%%%%%%%%%%%%%%%%%%%%%%%%%%%%%%%%%%%%%%%%%%%%%%%%%%%%%%%%%%%%%%%%%%%%%%%%%%%%%%%%%%%%%%%%%%%%%%%%%%%%%%%%%%%%%%%%

\section{Black hole evaporation}\label{sec:evap}
In Sec.\,\ref{sec:thermq}, we relied on the fact that a modified dispersion relation is expected to deform the Stefan-Boltzmann law in the trans-Planckian regime due to modifications on the relation between the energy density of the radiation fluid and its temperature. Therefore, a natural investigation that also deserves to be conducted concerns the effect that the thermal dimension has on the black hole evaporation rate.
\par
When one can decouple the thermal and athermal components of the black hole evaporation rate, the thermal contribution behaves according to the Stefan-Boltzmann law and the charge loss is governed by the Schwinger process:
\be \label{massloss}
\frac{dM}{dt}=-\frac{\pi^2}{15}\alpha\, \sigma\, T^4 +\frac{Q}{r_+}\frac{dQ}{dt},
\ee
where $\alpha$ is a constant related to the number of species of massless particles being emitted (since it is of order one, we shall set it to $1$ for simplicity) and $\sigma$ is the cross section of the black hole \cite{Hiscock:1990ex}. In this section, we shall discard the charge loss, since a deeper analysis of the impact of MDRs on the Schwinger effect would need to be carried out and it goes beyond the scope of this letter, therefore we only consider the Stefan-Boltzmann law contribution .
\par
Now, our goal is to analyze the first term on the right hand side of Eq.\,(\ref{massloss}) in the light of our MDR and its possible effects on the energy density and on the trajectories of massless particles in the black hole spacetime. 
\par
From the usual calculations of black body radiation with energy density $\rho$ \cite{Cardoso:2005cd}, the energy emitted by an infinitesimal element of area $da$ in the time interval $dt$ occupies a 2-dimensional hemisphere with radius $dr$ centered about $da$ in a direction $\theta$ and reads 
\be\label{du1}
dU_T=\rho(T)\, dr\, da\, \cos \theta.
\ee
We are setting the geometrical dimension of the spacetime to four (therefore we have the usual four-dimensional area and solid angle), and the thermal dimension manifests itself in the functional form of $\rho(T)$. From the above expression we see a further correction that should be considered: the MDR shall alter the relation between the time $dt$ elapsed when the radiation travel a distance $dr$. From the MDR (\ref{MDR}), we can derive the group velocity of the electromagnetic radiation and use it to substitute $dr\rightarrow  \partial E/\partial p|_{\Omega=0}\, dt$ in (\ref{du1}).
\par
To continue this calculation, we shall choose a specific dispersion relation in order to relate these quantities. For simplicity, let us consider $\ell_t=0$ in (\ref{MDR}), such that we shall keep corrections only due to powers of momentum implying that
\be
dr\rightarrow \frac{p+(1+\gamma_x)\ell_xp^{2\gamma_x+1}}{E}dt,\label{dr1}
\ee
which can be further simplified when we express the momentum in terms of the energy, $p=p(E)$.
\par
Thus, we can integrate (\ref{du1}) in order the find
\be\label{du2}
\frac{dU_T}{dt}\propto \rho(T)\, \frac{p(E)+(1+\gamma_x)\ell_xp(E)^{2\gamma_x+1}}{E}a,
\ee
where $a$ is the total area of the emitting surface, i.e., the geometrical optics cross section of the black body (which is not the event horizon) that is determined by the impact parameter of the metric.
\par
Since we are assuming a dispersion relation that is modified by a term proportional to the spatial momentum, in order to absorb the effect of the MDR not only on the equation of state but also on the trajectories of the fluid particles, we should consider a modification of the equation of motion of free particles along the equatorial plane. As the MDR reads $E^2-p^2-\ell_x^{2\gamma_x}p^{2(1+\gamma_x)}=0$, in order to derive a manageable expression and still capture essential properties of Planck-scale corrections, we assume a curved MDR of the form $p_0^2-p_r^2-r^2\dot{\phi}^2-\ell_x^{2\gamma_x}p_r^{2(1+\gamma_x)}=0$, where $p_0^2\doteq A(r)\dot{t}^2$, $p_r^2\doteq A(r)^{-1}\dot{r}^2$ are the squared time and radial components of the 4-momenta in this black hole metric (and ``dot'' means derivative with respect to an affine parameter $\lambda$). In this way, our proposed {\em ansatz} for the trajectories of photons in this curved spacetime reads
\be\label{eom}
A(r)\dot{t}^{\, 2}-A(r)^{-1}\dot{r}^{\, 2}-\ell_x^{2\gamma_x}A(r)^{-1-\gamma_x}\dot{r}^{\, 2(1+\gamma_x)}-r^2\dot{\phi}^{\, 2}=0,
\ee
where $A(r)$ is given by (\ref{metric}). Some quantities remain the same as if we were considering the trajectory of a massless particle obeying an undeformed dispersion relation, like the constraint obeyed by constant radii and the position of maxima of the potential of these world lines. In this way, we can define the classic impact parameter as usual (see Box 4.1 of Ref.\cite{livro1}):
\be
b_c=\frac{J}{E_{\infty}}=\frac{r_{\text{min}}^{3/2}}{\sqrt{r_{\text{min}}-2GM+Q(Q/r_{\text{min}})^{1+\gamma_x}}},
\ee
where $J\doteq r^2 \dot{\phi}$ and $E_{\infty}\doteq A(r)\dot{t}$ are the conserved angular momentum and energy, and $r_{\text{min}}$ is the radius of closest approach of massless particles, which is the critical point of the potential $V(r)=A(r)(J/r)^2$.
\par
However, we should notice that $b_c$ is not the physical impact parameter, since its definition assumes the usual dispersion relation in order to replace the momentum by the energy $E_{\infty}$. In fact, the square of the impact parameter is fundamentally given by $J^2/p^2$, which for our purposes can be cast in a more appropriate form from the MDR $E_{\infty}^2=p^2\left[1+\ell_x^{2\gamma_x}p^{2\gamma_x}\right]$. This observation leads us to derive the following correction:%\footnote{As a matter of fact, this correction has little impact on the behavior of the curve $dM(t)/dt$ as a function of $M$. However, we add it for completeness.}
\be\label{b2}
b^2_{\gamma_x}=b_c^2\left[1+\ell_x^{2\gamma_x}p(E_{\infty})^{2\gamma_x}\right],
\ee
where, as said before, $b_c=J/E_{\infty}$ is the classic impact parameter. This leads to an energy-dependent parameter, meaning that higher energetic photons probe larger surfaces defined by the impact parameter. 
\par
In order to derive a deformed evaporation rate for the black hole, we can assume a common {\em ansatz} used in investigations of modified dispersion relations or generalized uncertainty principles in the thermodynamics of black holes \cite{Adler:2001vs} consisting in identifying the characteristic energy of the emitted photons with the temperature of the Hawking radiation. Therefore, identifying the internal energy of the black hole with its mass, we state the following deformed Stefan-Boltzmann law of black hole evaporation in the trans-Planckian regime
\be\label{Mprime}
\small{\frac{dM_{d}}{dt}\propto -b_c^2\left[1+\ell_x^{2\gamma_x}p(T)^{2\gamma_x}\right]\left[p(T)+\ell_x^{2\gamma_x}(1+\gamma_x)p(T)^{2\gamma_x+1}\right]T^{3/(1+\gamma_x)}.}
\ee 
The first term in square brackets comes from the correction to the impact parameter given by Eq.(\ref{b2}), while the second one comes from the correction of the black body radiation law from Eq.(\ref{dr1}). When $\gamma_x=0=\ell_x$ (which implies that $p(T)\propto T$), we recover the usual law $dM/dt\propto T^4$. We stress that since these expressions are formulated in the high-energy regime by resorting to Eqs.\ (\ref{rho1}) and (\ref{dT1}), it is essential to perform this procedure in order to regain the undeformed case.
\par
To illustrate this method we shall study the case of {\it dimensional reduction}, which is reported in several approaches to the quantum gravity problem (see, for instance \cite{Amelino-Camelia:2013tla} and references therein). To simulate this effect, we consider $\gamma_x=2$ (furnishing thermal dimension 2 at the Planck scale). For simplicity, let us assume the parameter $\ell_x$ as equal to the Planck length $L_{\text{P}}=\sqrt{G}$.
\par
\begin{figure}[h]
\includegraphics[scale=.95]{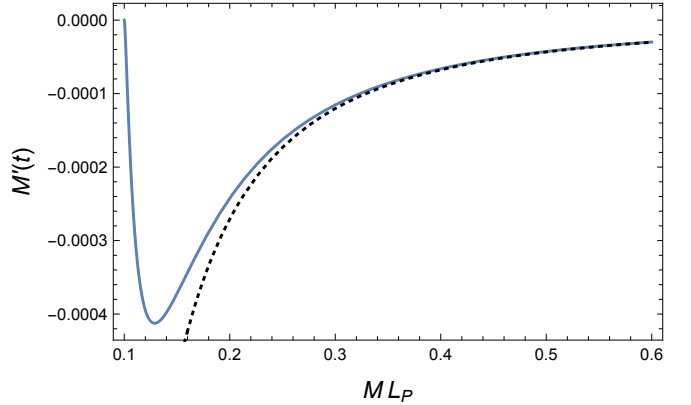}
\caption{$M'(t)$ as a function of $\ell_{\text{P}}M$ when $d=4$, $\ell_x=0$. Blue (solid) curve corresponds to the charged case ($Q/L_{\text{P}}=0.1$), and black (dashed) curve describes the Schwarzschild case.}\label{m1}
\end{figure}
\begin{figure}[h]
\includegraphics[scale=.95]{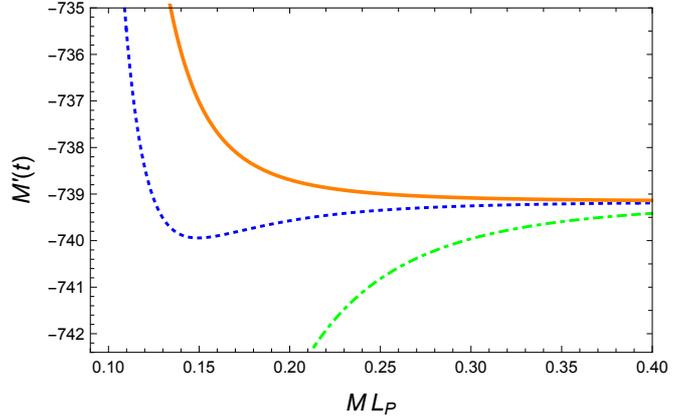}
\caption{$M'(t)$ as a function of $\ell_{\text{P}}M$ when $d=2$ and $\ell_x=L_{\text{P}}$. Orange (solid) curve corresponds to the charged case ($Q/L_{\text{P}}=0.076$), blue (dotted) curve corresponds to $Q/L_{\text{P}}=0.072$ and green (dash-dotted) curve describes the Schwarzschild case.}\label{m2}
\end{figure}

Picking the real root of the momentum in the on-shell relation, we replace the energy by the temperature, which can then be written as function of the mass $M$ and the charge $Q$ using Eq.(\ref{TH1}) and the outer horizon $r_+=r_+(M,Q)$ that is found from the solution of $A(r_+)=0$. Thus, we are able to depict the behavior of the evaporation rate of the black hole as a function of its mass (for fixed values of the charge) in Figs.\,(\ref{m1}) and (\ref{m2}) in units of Planck length.

The blue (solid) curve of Fig.(\ref{m1}) describes the usual Reissner-Nordstr\"{o}m case when $d=4$, where we verify that the absolute value of $dM(t)/dt$ increases as the mass decreases in a way that is similar to the Schwarzschild case (black, dotted curve). However, since we are not considering the loss of massive charged particles, the Reissner-Nordstr\"{o}m black hole cannot lose mass indefinitely, so it must reach the extremal case where the temperature vanishes. Therefore, we have a critical value for the mass where the curve turns up until it reaches $dM(t_f)/dt=0$ when $M=Q/G$, while in the uncharged case the decay continues until it reaches the catastrophic evaporation where the rate of mass loss diverges.
\par
Considering now the effect of dimensional reduction ($d=2$) in Fig.\,(\ref{m2}) (green, dash-dotted curve), since the MDR does not alter the solution of the uncharged black hole in Eq.\ (\ref{metric1}) and consequently the temperature as function of $M$ and $Q$ (implicitly determined by $r_+$ in Eq.(\ref{TH1})), the corrections that we introduce in the Stefan-Boltzmann law (relating $M'(t)$ and $T$) are not enough to cause a qualitative modification on the shape of the mass loss rate.\footnote{Maybe in an alternative scenario, where one considers a different modified dispersion relation, for instance, a non-polynomial form or with negative polynomial factors, one might derive major differences depending on the phenomenological interest.} 
\par
On the other hand, the curve of the dimensionally reduced Reissner-Nordstr\"{o}m evaporation rate behaves like the $4D$ case for low charges (see the blue-dotted curve of Fig.\,(\ref{m2})), but gets significantly modified when considering higher charges by presenting a convex nature, i.e., always going upwards, without mimicking the uncharged case for high masses (orange, solid line). This property is a clear signature of dimensional reduction in the charged case due to the modification that $\omega$ induces on the metric. It is found numerically that this transition to the convex behavior happens at a fixed value of charge $Q/L_{\text{P}} \approx 0.075$.
\par
In fact, when $d=4$, the quantity $M'(t)$ as a function of $M$ presents the following critical point
\be
(M_4)_{c}=\sqrt{\frac{19+3\sqrt{2}}{14}}\frac{Q}{G},
\ee
which means that the undeformed case always has a critical mass for any non-zero charge.
\par
Also notice that the scale of mass loss rate is six orders of magnitude higher in the dimensionally reduced case when compared to the usual one. This is due to the specific form of the function $\rho(T)$ in the Planckian temperature scale.

%%%%%%%%%%%%%%%%%%%%%%%%%%%%%%%%%%%%%%%%%%%%%%%%%%%%%%%%%%%%%%%%%%%%%%%%%%%%%%%%%%%%%%%%%%%%%%%%%%%%%%%%%%%%%%%%%%%%%%%%%%%%%%%%%%%%%%%%%%%%%%%%%%%%%%%%%%%

\subsection{Asymptotic behavior}
From the preceding section, we analyze the asymptotic behavior of the modified Reissner-Nordstr\"{o}m and Schwarzschild black holes and verify some interesting similarities with the undeformed cases. In fact, the charged case tends to the extremal black hole as the mass diminishes and approaches $M\rightarrow (16/27)^{1/4}Q$, where $T\rightarrow 0$. In this limit, we shall regain the usual undeformed behavior, since departures from the modified dispersion relation are manifest only in a regime of high temperatures of the surrounding radiation fluid of the black hole. The low temperature regime is characterized by presenting the radiation equation of state parameter $\omega= 1/3$, which is described by adopting $\ell_x\rightarrow 0$. Therefore assuming the above requirements ($T\rightarrow 0$ and $\ell_x\rightarrow 0$), we shall derive the asymptotic behavior
\be
\left|\frac{1}{b_c^2}\frac{dM_2}{dt}\right|\propto T^4.
\ee
\par
On the other hand, for the Schwarzschild metric there is no critical mass (where $M'(t)=0$), which means that as the black hole mass decreases, the temperature raises indefinitely, suggesting a completely different scenario in which the characteristic energies and momenta involved in the process get higher and higher, indicating that we should discard squared momenta terms in comparison to those of sixth order in the MDR, i.e., we should have $E^2\propto p^6$. This observation also indicates that one should consider just terms proportional to $\ell_x^4$ in Eq.\ (\ref{Mprime}). Besides, as we identified the characteristic energy of the emitted photons with the Hawking temperature, we shall have $p\propto T^{1/3}$. Curiously, this also leads to an asymptotic behavior of the form
\be
\left|\frac{1}{b_c^2}\frac{dM_2}{dt}\right|_{Q=0} \propto T^4, 
\ee
which implies a restoration of the original behavior of the mass loss process in four dimensions, showing an identification between infrared and ultraviolet regimes.

%%%%%%%%%%%%%%%%%%%%%%%%%%%%%%%%%%%%%%%%%%%%%%%%%%%%%%%%%%%%%%%%%%%%%%%%%%%%%%%%%%%%%%%%%%%%%%%%%%%%%%%%%%%%%%%%%%%%%%%%%%%%%%%%%%%%%%%%%%%%%%%%%%%%%%%%%%%

\section{Concluding remarks}\label{sec:conc}
The assumption that photons obey a modified dispersion relation as consequence of a quantum gravity regime leaves imprints on the equation of state parameter of the radiation fluid. In the ultraviolet limit, such deformed quantity approaches a constant (in general, different from the usual $1/3$ value), which allows the introduction of the notion of a thermal dimension that follows from a modified Stefan-Boltzmann law. In this paper, we report the influence that these results have on the metric and on the thermodynamics of a black hole surrounded by an imperfect radiation fluid (for which $\omega$ relates the average pressure with the energy density) derived from Kiselev's approach in order to describe a deformed Reissner-Nordstr\"{o}m spacetime.
\par
As a direct application of these results, we analyzed the evaporation rate of the black hole due to the loss of massless particles from the point of view of the aforementioned Stefan-Boltzmann law and also considering kinematical effects. Thus, we managed to study not only the charged, but also the uncharged case when the thermal dimension becomes two. Regarding the former, we found the usual behavior towards an extremal black hole, but without the usual critical mass. Instead, for highly charged black holes we have an ever convex curve (recovering the usual qualitative behavior only for smaller charges). The uncharged black hole qualitatively behaves in a similar way to the undeformed case, independently of the thermal dimension. It should be stressed that the scale of the evaporation rate of the four and two thermal dimension cases differs by six orders of magnitude due to their different relations between the energy density and the temperature, which consists in the main motivation of this work. Such observation may have important phenomenological consequences and were partly explored in a perturbative study when analyzing Planck-scale modified dispersion relations and generalized uncertainty principles in \cite{Amelino:2006}, but was not extended to this higher energy regime. We also computed the general asymptotic behavior of the dimensionally reduced case, indicating a return to the original four-dimensional behavior in the final stages of evaporation of the Reissner-Nordstr\"{o}m and Schwarzschild black holes.
\par
An interesting analysis that should be carried out in the future concerns the loss of massive charged particles, which would require a more detailed investigation regarding the Schwinger effect that governs this process. Another exciting exploration regards the effect that primordial charged black holes have on dark matter \cite{Lehmann:2019zgt}, a subject that shall be investigated in the future since imprints from a quantum gravitational regime may still be present in such primordial objects.

%%%%%%%%%%%%%%%%%%%%%%%%%%%%%%%%%%%%%%%%%%%%%%%%%%%%%%%%%%%%%%%%%%%%%%%%%%%%%%%%%%%%%%%%%%%%%%%%%%%%%%%%%%%%%%%%%%%%%%%%%%%%%%%%%%%%%%%%%%%%%%%%%%%%%%%%%%%

\section*{Acknowledgements}
IPL would like to acknowledge the contribution of the COST Action CA18108. The work of IPL was partially supported by the National Council for Scientific and Technological Development - CNPq grant 306414/2020-1. GBS is financially supported by CAPES through the grant 88882.317979/2019-1. The authors would like to thank Giulia Gubitosi for her useful comments on a previous version of this manuscript. 

%The authors would like to thank CNPq (Conselho Nacional de Desenvolvimento Cient\'ifico e Tecnol\'ogico, Brazil) for financial support. 

%%%%%%%%%%%%%%%%%%%%%%%%%%%%%%%%%%%%%%%%%%%%%%%%%%%%%%%%%%%%%%%%%%%%%%%%%%%%%%%%%%%%%%%%%%%%%%%%%%%%%%%%%%%%%%%%%%%%%%%%%%%%%%%%%%%%%%%%%%%%%%%%%%%%%%%%%%%

%\bibliography{refs}
\end{document}